\documentclass[pra,twocolumn,amsmath,amssymb,superscriptaddress,
showpacs]{revtex4}

\usepackage{graphics}
\usepackage{epsfig}
\usepackage[latin1]{inputenc} % Umlaute!
\usepackage{pstricks,pst-grad,fancybox,graphics}
\usepackage{amsmath}
\usepackage{amssymb}
\usepackage{enumerate}

\newcommand{\be}{\begin{equation}}
\newcommand{\ee}{\end{equation}}
\newcommand{\BE}{\begin{equation}}
\newcommand{\EE}{\end{equation}}
\newcommand{\eea}{\end{eqnarray}}
\newcommand{\bea}{\begin{eqnarray}}

\newcommand{\mean}[1]{\ensuremath{\langle{#1}\rangle}}

\newcommand{\eins}{\openone}

\newcommand{\BB}{\ensuremath{\mathcal{B}}}
\newcommand{\NN}{\ensuremath{\mathcal{N}}}

\newcommand{\ket}[1]{\ensuremath{|#1\rangle}}

\newcommand{\kommentar}[1]{}
\newcommand{\meanlhv}[1]{\ensuremath{\langle{#1}\rangle_{\rm LHV}}}

\begin{document}
\title{Generalized Ardehali-Bell inequalities for graph states}
\author{Otfried G{\"u}hne}
\affiliation{Institut f\"ur Quantenoptik und Quanteninformation,
\"Osterreichische Akademie der Wissenschaften, A-6020 Innsbruck,
Austria}
\affiliation{Institut f\"ur Theoretische Physik,
Universit\"at Innsbruck, A-6020 Innsbruck,
Austria}
\email{otfried.guehne@uibk.ac.at}
\author{Ad\'an Cabello}
\affiliation{Departemento de F\'{\i}sica Aplicada II, Universidad de
Sevilla, E-41012 Sevilla, Spain}
\email{adan@us.es}
\date{\today}

%First version: August 2007.
%This version: October 27, 2007.

%%%%%%%%%%%%%%%%%%%%%%%%%%%%%%%%%%%%%%%%%%%%%%%%%%%%%%%%%%%%%%%%%%%

\begin{abstract}
We derive Bell inequalities for graph states by generalizing the
approach proposed by Ardehali [Phys. Rev. A {\bf 46}, 5375 (1992)]
for Greenberger-Horne-Zeilinger (GHZ) states. Using this method, we
demonstrate that Bell inequalities with nonstabilizer observables
are often superior to the optimal GHZ-Mermin-type (or
stabilizer-type) Bell inequalities.
\end{abstract}

%%%%%%%%%%%%%%%%%%%%%%%%%%%%%%%%%%%%%%%%%%%%%%%%%%%%%%%%%%%%%%%%%%%

%PACS: bell ineq, quant comp, error correction

\pacs{03.65.Ud, 03.67.Lx, 03.67.Pp}

\maketitle

%%%%%%%%%%%%%%%%%%%%%%%%%%%%%%%%%%%%%%%%%%%%%%%%%%%%%%%%%%%%%%%%%%%%%

\section{Introduction}
\label{Sec1}

%%%%%%%%%%%%%%%%%%%%%%%%%%%%%%%%%%%%%%%%%%%%%%%%%%%%%%%%%%%%%%%%%%%%%

Bell inequalities are constraints imposed by local hidden variable
(LHV) models on the correlations of distant experiments. The fact
that quantum mechanics predicts a violation of these relations makes
them useful for demonstrating the impossibility of LHV models. In
addition, Bell inequalities are useful as entanglement witnesses
\cite{HGBL05}, and can be used for demonstrating the security of
some quantum key distribution protocols \cite{Ekert91, ABGMPS07}.

The set of all LHV models corresponds to a polytope in a
high-dimensional space of correlations, and Bell inequalities
correspond to its facets \cite{Peres99}. The classification of this
set is the subject of intensive research, and a complete
classification has been achieved only for some specific cases
\cite{BelinskiiKlyshko93, WW01, ZB02}.

However, given a quantum state, it is not clear which Bell
inequality is the one maximally violated by this state, because the
above-mentioned results do not allow us to specify the optimal
measurement observables in a simple way. This specification,
however, is important for any experiment. Moreover, finding Bell
inequalities with a high amount of violation for a given state
allows us to investigate the interplay between the violation of
local realism and decoherence.

Greenberger, Horne, and Zeilinger (GHZ) showed that multipartite
states which are simultaneous eigenstates of several local
observables (henceforth called GHZ states) can lead to striking
contradictions with local realism if we consider the perfect
correlations between these local observables \cite{GHZ89}. The idea
to construct Bell inequalities from such perfect correlations has
been extended by Mermin and others into several directions
\cite{Mermin90, GTHB05, SASA05, Cabello05, TGB06, CGR07, CM07, DP97,
Hsu06}. The extensions include GHZ states with more particles,
cluster states, and graph states---which generalize GHZ states and
are of great importance for many applications in quantum information
\cite{HDERVB06}.

Remarkably, as early as in 1992, Ardehali showed that for special
examples of GHZ states, other Bell inequalities exist which lead to
a higher violation of local realism compared to the GHZ-Mermin-type
(or stabilizer-type) inequalities \cite{Ardehali92}.

In this paper we generalize Ardehali's method to arbitrary graph
states. This allows us to derive Bell inequalities with a high
amount of violation for a variety of different states. We find that
the inequalities are often superior to the optimal Bell inequalities
of the GHZ-Mermin-type, suggesting that the curious property
discovered by Ardehali is quite generic. Interestingly, although the
Ardehali approach was originally designed to derive Bell
inequalities which use not only perfect correlations, our extended
method also allows us to derive some Bell inequalities of the
stabilizer-type for tree graph states (i.e., graph states associated
to graphs which do not contain any closed loop).

This paper is organized as follows: In Sec.~\ref{Sec2} we give a
short definition of graph states. In Sec.~\ref{Sec3} we reformulate
Ardehali's method in the language of stabilizing operators and graph
states. Then, in Sec.~\ref{Sec4} we apply the method to obtain Bell
inequalities for five- and six-qubit graph states, showing that the
Ardehali approach delivers higher violations of local realism than
GHZ-Mermin-type inequalities. In Sec.~\ref{Sec5} we further extend
the method and derive some general Bell inequalities for tree graph
states. Finally, in Sec.~\ref{Sec6} we sum up the results.

%%%%%%%%%%%%%%%%%%%%%%%%%%%%%%%%%%%%%%%%%%%%%%%%%%%%%%%%%%%%%%%%%%%%%

%Fig. 1

%%%%%%%%%%%%%%%%%%%%%%%%%%%%%%%%%%%%%%%%%%%%%%%%%%%%%%%%%%%%%%%%%%%%%

\begin{figure}[h]
\centerline{\includegraphics[width=0.90\columnwidth]{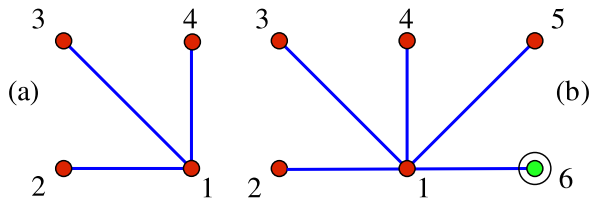}}
\caption{\label{ghzbild} (Color online) The star graph for (a) four
and (b) six vertices. The corresponding graph state is, up to local
unitary transformations, the GHZ state. The Ardehali inequality for
the six-qubit GHZ state can be derived from the Mermin inequality
for the five-qubit GHZ state, when a sixth qubit is added. See
Sec.~\ref{Sec3} for details.}
\end{figure}

%%%%%%%%%%%%%%%%%%%%%%%%%%%%%%%%%%%%%%%%%%%%%%%%%%%%%%%%%%%%%%%%%%%%%

\section{Graph states}
\label{Sec2}

%%%%%%%%%%%%%%%%%%%%%%%%%%%%%%%%%%%%%%%%%%%%%%%%%%%%%%%%%%%%%%%%%%%%%

Graph states are a family of multi-qubit states that play a crucial
role in many applications of quantum information theory, such as
quantum error correcting codes, measurement-based quantum
computation, and quantum simulation \cite{HDERVB06, HRD07}.
Consequently, a significant experimental effort is devoted to the
creation and investigation of graph states \cite{WRRSWVAZ05,
KSWGTUW05, LZGGZYGYP07, LGGZCP07, PWSKPW07}. Graph states are
defined as follows:

Let $G$ be a graph, i.e., a set of $N$ vertices corresponding to
qubits and edges connecting them. Some examples are shown in
Figs.~\ref{ghzbild}--\ref{sechsbild}. For each vertex $i$, the
neighborhood $\NN(i)$ denotes the vertices which are connected with
$i.$ Then, we can associate a stabilizing operator $g_i$ to each
vertex $i$ by
\begin{equation}
g_i:= X^{(i)}\bigotimes\nolimits_{j\in \NN(i)} Z^{(j)}.
\label{stabdef}
\end{equation}
Here and in the following, $X^{(i)}, Y^{(i)}, Z^{(i)}$ denote the
Pauli matrices $\sigma_x,\sigma_y,\sigma_z,$ acting on the $i$th
qubit. The index ${(i)}$ may be omitted where there is no risk of
confusion. The graph state $\ket{G}$ associated with the graph $G$
is the unique $N$-qubit state fulfilling
\begin{equation}
g_i \ket{G}= \ket{G}, \mbox{ for } i=1,\ldots,n. \label{graphdef}
\end{equation}
Physically, the stabilizing operators describe the perfect
correlations in the state $\ket{G},$ since
$\mean{g_i}=\mean{X^{(i)}\bigotimes_{j\in N(i)} Z^{(j)}}=1.$ In the
GHZ-Mermin-type (or stabilizer-type) Bell inequalities, these
perfect correlations are used to derive contradictions to local
realism \cite{GHZ89, Mermin90, GTHB05, SASA05, TGB06, Cabello05,
CGR07}.

For our later discussion, it is important to note that different
graphs may lead to graph states which differ only by a local unitary
transformation; that is, their entanglement properties are the same.
The main graph transformation, which leaves the entanglement
properties invariant, is the so-called local complementation
\cite{HDERVB06}.

Local complementation acts on a graph as follows: one picks out a
vertex $i$ and inverts the neighborhood $\NN(i)$ of $i$; that is,
vertices in the neighborhood which were connected become
disconnected and vice versa. It has been shown that local
complementation acts on the graph state as a local unitary
transformation of the Clifford type, and therefore leaves the
nonlocal properties invariant. More specifically, it induces the map
$Y^{(i)} \mapsto Z^{(i)}, Z^{(i)} \mapsto -Y^{(i)}$ on the qubit
$i$, and the map $X^{(j)} \mapsto -Y^{(j)}, Y^{(j)} \mapsto X^{(j)}$
on the qubits $j\in \NN(i)$ of the neighborhood \cite{HDERVB06}. On
the level of the generators, it maps the generators $g^{\rm{old}}_j$
with $j\in \NN(i)$ to $g^{\rm{new}}_j g^{\rm{new}}_{i}.$ We will see
later an explicit example of this transformation. Finally, it should
be noted that not all local unitary transformations between graph
states can be represented by a local complementation. An
example has been recently found \cite{JCWY07}.

%%%%%%%%%%%%%%%%%%%%%%%%%%%%%%%%%%%%%%%%%%%%%%%%%%%%%%%%%%%%%%%%%%%%%

\section{The basic method}
\label{Sec3}

%%%%%%%%%%%%%%%%%%%%%%%%%%%%%%%%%%%%%%%%%%%%%%%%%%%%%%%%%%%%%%%%%%%%%

In order to derive our Bell inequalities, let us first reformulate
the original Ardehali method \cite{Ardehali92} (see also
\cite{CAF06}) in the language of stabilizing operators and graph
states. He considers an $N$-qubit GHZ state
\begin{equation}
\ket{{\rm GHZ}}=\frac{1}{\sqrt{2}}(\ket{00\ldots0}+\ket{11\ldots1}),
\label{GHZdef}
\end{equation}
which is an eigenstate with eigenvalue one of the stabilizing
operators:
\begin{subequations}
\begin{align}
&g_1=X^{(1)}X^{(2)}\ldots X^{(N)}, \\
&g_2 =Z^{(1)}Z^{(2)}\openone^{(3)}\ldots \openone^{(N)}, \\
&g_3 = Z^{(1)}\openone^{(2)}Z^{(3)}\ldots \openone^{(N)},\ldots,\\
&g_N = Z^{(1)}\openone^{(2)}\ldots \openone^{(N-1)}Z^{(N)}.
\end{align}
\end{subequations}
These are, up to a local change of the basis, stabilizing operators
like those in Eq.~(\ref{stabdef}). The GHZ state corresponds to a
star graph like those shown in Fig.~\ref{ghzbild}.

A Bell inequality for the GHZ state is the Mermin inequality
\cite{Mermin90}, which has the Bell operator
\begin{equation}
\BB_N = g_1 \prod_{k=2}^N (\openone + g_k). \label{MerminBellop}
\end{equation}
According to quantum mechanics, the expectation value of $\BB_N$ for
the GHZ state is $\mean{\BB_N} = 2^{N-1}$, while for local realistic
models $\meanlhv{\BB_B}\leq C_N,$ with $C_{N} = {2}^{(N-1)/2}$
($C_{N} = {2}^{N/2}$) for $N$ odd (even). The amount of violation is
then $V_N=\mean{\BB_N}/C_{N} ={2}^{(N-1)/2}$ ($V_N={2}^{(N-2)/2})$
for $N$ odd (even).

We can rewrite the Bell operator (\ref{MerminBellop}) as
\begin{subequations}
\begin{align}
&\BB_N = \left(g_1 \prod_{k=2}^{N-1} (\openone + g_k)\right)
(\openone
+ g_N) \\
&= \BB_{N-1} \otimes X^{(N)} + \tilde{\BB}_{N-1} \otimes Y^{(N)} \label{crucial} \\
&= \frac{1}{\sqrt{2}} \big[ \BB_{N-1} \otimes (A^{(N)}+B^{(N)}) +
\tilde{\BB}_{N-1}\otimes(A^{(N)}-B^{(N)}) \big], \label{6c}
\end{align}
\end{subequations}
where $A^{(N)}=(X^{(N)}+Y^{(N)})/\sqrt{2}$ and
$B^{(N)}=(X^{(N)}-Y^{(N)})/\sqrt{2}$, and $\tilde{\BB}_{N-1}$
denotes the Bell operator which is obtained from ${\BB}_{N-1}$ by
making a transformation $X^{(1)} \mapsto - Y^{(1)}; Y^{(1)}\mapsto
X^{(1)}$ on the first qubit.

We take the right hand side of Eq.~(\ref{6c}) as a new Bell
operator $\BB_N^{({\rm Ardehali})}$, with two new measurement
directions $A^{(N)}$ and $B^{(N)}$ on qubit $N.$ For LHV models,
we have
\begin{align}
&
\meanlhv{
\BB_{N-1}\otimes(A^{(N)}+B^{(N)})
+ \tilde{\BB}_{N-1}\otimes (A^{(N)}-B^{(N)})
}
\nonumber
\\
& \;\;\;\;\;\; \leq 2 \max\big[ \sup_{\rm {LHV}}\meanlhv{ \BB_{N-1}
}, \sup_{\rm {LHV}}\meanlhv{\tilde{\BB}_{N-1}} \big],
\label{trick1}
\end{align}
so we have for the Bell operator $\BB_N^{({\rm Ardehali})}$
\begin{equation}
\sup_{\rm {LHV}}\meanlhv{\BB_N^{({\rm Ardehali})}} = \sqrt{2} \sup_{\rm
{LHV}}\meanlhv{\BB_{N-1}}.
\end{equation}
Since $\BB_N$ and $\BB_N^{({\rm Ardehali})}$ are the same if considered
as operators acting on a Hilbert space, the maximum value for quantum
states coincides and equals $2^{N-1}$. Therefore, we obtain a violation
of local realism by a factor of
\begin{equation}
\tilde{V}_N^{({\rm Ardehali})} =
\frac{2^{N-1}}{\sqrt{2} \sup_{\rm {LHV}}\meanlhv{\BB_{N-1}}}
=
\sqrt{2} V_{N-1},
\end{equation}
which is larger than $V_N$, if $N$ is even.

One can interpret this method as follows: We start with an
$(N-1)$-qubit GHZ state described by a star graph, and add a qubit
[see Fig.~\ref{ghzbild}(b)]. Then, we consider the Bell operator
$\BB_{(N-1)}$ as an extended operator $\BB_{(N-1)}^{\rm (Ext)}$ on
all $N$ qubits, and multiply it by $(\openone + g_N)$. Making a
replacement $X^{(N)}/Y^{(N)}\rightarrow (A^{(N)}\pm B^{(N)})$ on the
$N$th qubit, we obtain a Bell inequality which is violated by the
$N$-qubit state by an amount of $\tilde{V}_N = \sqrt{2} V_{N-1}.$

\begin{figure}[t]
\centerline{\includegraphics[width=0.9\columnwidth]{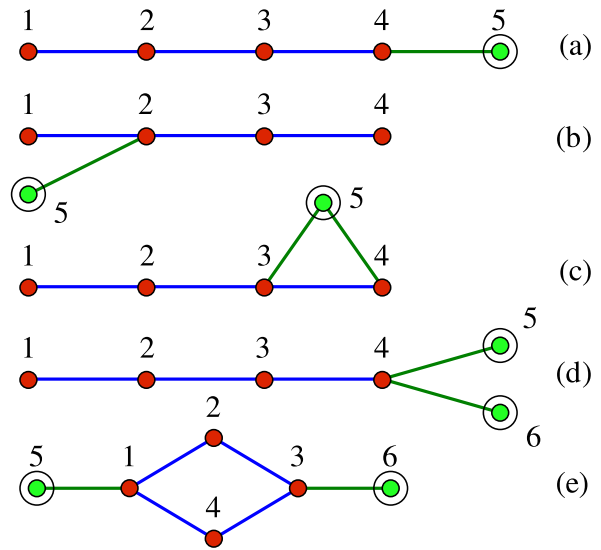}}
\caption{\label{clusterbild} (Color online) Different possibilities
to derive Bell inequalities for states on five or more qubits. The
graph of the first four qubits corresponds to the four-qubit cluster
state. See Sec.~\ref{Sec4} for details.}
\end{figure}

Under which conditions does this method work? A first condition can
be seen from Eq.~(\ref{crucial}): There, it is required that in the
first term (where we multiply $\BB_{N-1}^{\rm (Ext)}$ with
$\openone$, not $g_N$) we have a nontrivial observable (here
$X^{(N)}$) on the added $N$th qubit, and not $\openone^{(N)}.$ In
our example, this stemmed from the fact that the stabilizer element
$g_1$ is a factor in all terms of the Bell operator $\BB_{N-1}^{\rm
(Ext)}.$ This condition may also be fulfilled in other cases. For
instance, if $\BB_{N-1}^{\rm (Ext)}$ contains a factor $(g_k +g_l),$
one may add a qubit and connect it (in the graph state sense)
directly to both the qubits $k$ and $l.$ We will see an example
later [see Fig.~\ref{clusterbild}(c)].

A second condition comes from the fact that the multiplication of
$\BB_{N-1}^{\rm (Ext)}$ with $g_N$ must give again a Bell operator
(with the same bound for LHV models as $\BB_{N-1}$) on the first
$N-1$ qubits. In Eq.~(\ref{crucial}) this was fulfilled since the
multiplication with $g_N$ induced only a relabeling of the variables
in $\BB_{N-1}$.

Finally, it should be noted that in some cases we may add several
qubits consecutively. By adding several qubits one can derive a
similar bound as in Eq.~(\ref{trick1}), and one can then directly
obtain a Bell inequality with a violation of $V_N= (\sqrt{2})^k
V_{N-k}.$

%%%%%%%%%%%%%%%%%%%%%%%%%%%%%%%%%%%%%%%%%%%%%%%%%%%%%%%%%%%%%%%%%%%%%

%Fig. 2

%%%%%%%%%%%%%%%%%%%%%%%%%%%%%%%%%%%%%%%%%%%%%%%%%%%%%%%%%%%%%%%%%%%%%

%%%%%%%%%%%%%%%%%%%%%%%%%%%%%%%%%%%%%%%%%%%%%%%%%%%%%%%%%%%%%%%%%%%%%

\section{Application to Bell inequalities for the
four-qubit cluster state} \label{Sec4}

%%%%%%%%%%%%%%%%%%%%%%%%%%%%%%%%%%%%%%%%%%%%%%%%%%%%%%%%%%%%%%%%%%%%%

The method presented in Sec.~\ref{Sec3} does not allow us to obtain
new Bell inequalities for four-qubit graph states: The only
three-qubit graph state is the GHZ state \cite{HDERVB06}, with the
Mermin inequality as the relevant Bell inequality. As can be easily
seen, the only way to derive an Ardehali-type Bell inequality for
four-qubit states from that results is the Ardehali inequality for
the four-qubit GHZ state, which is already known \cite{Ardehali92}.

However, we can apply the method to Bell inequalities for the
four-qubit cluster state and obtain Ardehali-type Bell inequalities
for different five-qubit graph states.

%%%%%%%%%%%%%%%%%%%%%%%%%%%%%%%%%%%%%%%%%%%%%%%%%%%%%%%%%%%%%%%%%%%%%

\subsection{Five-qubit states}

%%%%%%%%%%%%%%%%%%%%%%%%%%%%%%%%%%%%%%%%%%%%%%%%%%%%%%%%%%%%%%%%%%%%%

%\subsection{Five-qubit linear cluster state}

%%%%%%%%%%%%%%%%%%%%%%%%%%%%%%%%%%%%%%%%%%%%%%%%%%%%%%%%%%%%%%%%%%%%%

Let us start by applying the method to the Bell inequalities for the
four-qubit (linear) cluster state derived by Scarani {\em et
al.}~\cite{SASA05, TGB06, Cabello05, CGR07}. The graph of this state
is shown in Fig.~\ref{clusterbild}. One of the suitable inequalities
is given \cite{CGR07} by
\begin{subequations}
\begin{align}
\BB_1^{\rm (LC4)} &= (\openone + g_1)g_2 (\openone+g_3)g_4
\\
&= Z X \openone X - ZYXY + Y Y \openone X + YXXY. \label{sasa1}
\end{align}
\end{subequations}
Here, we omit the indices on the Pauli operators for simplicity.
This Bell inequality leads to a violation $V_4 = 2,$ since for the
cluster state $\mean{\BB_1^{\rm (LC4)}}=4,$ and the bound for LHV
models is 2. Since $g_4$ is a factor of the Bell operator, we can
add a qubit at the end [see Fig.~\ref{clusterbild}(a)] and obtain
the Bell operator
\begin{subequations}
\begin{align}
&\BB_1^{\rm (LC5)} =(\openone + g_1)g_2 (\openone+g_3)g_4
(\openone+g_5)
\\
&= \big[ (ZX\openone X - ZYXY + YY\openone X + YXXY) (A+B)
\nonumber \\
&+ (ZX\openone Y + ZYXX + YY\openone Y - YXXX) (A-B) \big]\big/
\sqrt{2},
\end{align}
\end{subequations}
with $A=(Z+Y)/\sqrt{2}$ and $B=(Z-Y)/\sqrt{2}$ \cite{graphremark}.
This Bell operator requires a measurement of 16 correlation terms.
It has a maximum value of $2\sqrt{2}$ for LHV models, leading to a
violation $V=2\sqrt{2}\approx 2.82$ for the five-qubit linear
cluster state. Remarkably, if only stabilizing operators are
considered, it can be proven that the maximal achievable violation
is only $5/2 = 2.5$ \cite{CGR07}. Therefore, for the five-qubit
linear cluster state, the violation of local realism can be
increased by considering Ardehali-type inequalities, as for the GHZ
state with a even number of qubits.

%%%%%%%%%%%%%%%%%%%%%%%%%%%%%%%%%%%%%%%%%%%%%%%%%%%%%%%%%%%%%%%%%%%%%

%\subsection{Five-qubit Y state}

%%%%%%%%%%%%%%%%%%%%%%%%%%%%%%%%%%%%%%%%%%%%%%%%%%%%%%%%%%%%%%%%%%%%%

Since $g_2$ is also a factor of $\BB_1^{\rm (LC4)}$, we can also
connect the fifth qubit to the second qubit [see
Fig.~\ref{clusterbild}(b)]. This leads to a Bell inequality for the
five-qubit Y-state,
\begin{align}
&\BB^{\rm (Y5)} = \big[ (ZX\openone X - ZYXY + YY\openone X + YXXY)
(A+B)
\nonumber \\
&+ (ZY\openone X + ZXXY - YX\openone X + YYXY) (A-B)
\big]\big/{\sqrt{2}},
\end{align}
with $A=(Z+Y)/\sqrt{2}$ and $B=(Z-Y)/\sqrt{2}$ \cite{graphremark}.
Therefore, the five-qubit Y-state also has a violation of
$V=2\sqrt{2}\approx 2.82.$ If only stabilizing operators are
considered, the maximal achievable violation is only $7/3 \approx
2.33$ \cite{CGR07}, proving again the superiority of the method
presented here.

%%%%%%%%%%%%%%%%%%%%%%%%%%%%%%%%%%%%%%%%%%%%%%%%%%%%%%%%%%%%%%%%%%%%%

%\subsection{Five-qubit linear cluster state (II)}

%%%%%%%%%%%%%%%%%%%%%%%%%%%%%%%%%%%%%%%%%%%%%%%%%%%%%%%%%%%%%%%%%%%%%

Finally, a further inequality is given by the Bell operator
$\BB_2^{\rm (LC4)} = (\openone+g_1)g_2 (g_3 + g_4)$ \cite{CGR07}.
Here, $(g_3 + g_4)$ is a factor of $\BB_2^{\rm (LC4)},$ hence we can
add a qubit connected to qubits $3$ and $4$ at the same time, as
shown in Fig.~\ref{clusterbild}(c). The resulting five-qubit state
also has a violation of $2\sqrt{2}.$ The Bell operator is given by
$\BB_1^{\rm (2c)} = (\openone+g_1)g_2 (g_3 + g_4)(\openone + g_5),$
where we have to introduce the measurements $A=(Z+Y)/\sqrt{2}$ and
$B=(Z-Y)/\sqrt{2}$ on the fifth qubit.

This state, however, was discussed before. Local complementation on
the vertex $4$ requires us to invert its neighborhood, consisting of
vertices $3$ and $5$, which become disconnected. Therefore, the
state in Fig.~\ref{clusterbild}(c) is equivalent to the five-qubit
linear cluster state in Fig.~\ref{clusterbild}(a). If we make the
local complementation, we obtain $\BB^{\rm (LC5)}=(\openone+g_1)g_2
(\openone + g_3)(g_4+ g_5)$, with the measurements
$A=(Z+X)/\sqrt{2}$ and $B=(Z-X)/\sqrt{2}$ on the fifth qubit, as a
Bell operator for the linear cluster state.

%%%%%%%%%%%%%%%%%%%%%%%%%%%%%%%%%%%%%%%%%%%%%%%%%%%%%%%%%%%%%%%%%%%%%

\subsection{Six-qubit states}

%%%%%%%%%%%%%%%%%%%%%%%%%%%%%%%%%%%%%%%%%%%%%%%%%%%%%%%%%%%%%%%%%%%%%

Let us now demonstrate how the method presented here can be used to
derive Bell inequalities for six-qubit states directly from the
four-qubit inequalities.

First, as already mentioned, one can add two qubits consecutively to
the cluster state and obtain the six-qubit Y state [see
Fig.~\ref{clusterbild}(d)]. Taking $\BB_1^{\rm (LC4)}$ as above,
considering $\BB^{\rm (Y6)} =(\openone + g_1)g_2 (\openone+g_3)g_4
(\openone+g_5) (\openone+g_6),$ and replacing $A=(Z+Y)/\sqrt{2}$ and
$B=(Z-Y)/\sqrt{2}$ on the fifth and sixth qubit, one finds that the
depicted six-qubit graph state has a violation of $V=4.$ Note that
the state in Fig.~\ref{clusterbild}(d) is of special interest in
quantum information science, since it allows a demonstration of
anyonic statistics in the Kitaev model \cite{HRD07, LGGZCP07,
PWSKPW07}.

Another interesting example is shown in Fig.~\ref{clusterbild}(e).
First, by a suitable sequence of local complementations and
relabeling of the qubits (first one makes a local complementation on
the second qubit, then on the third, and finally again on the
second), one can transform the graph of a four-qubit cluster state
in the form of a ``box,'' as shown here. Afterwards, the Bell
inequality in Eq.~(\ref{sasa1}) reads
\begin{equation}
\BB^{\rm(Box4)} = g_1 (\eins + g_2) (\eins + g_4).
\end{equation}
We can add a fifth qubit as in Fig.~\ref{clusterbild}(e) and achieve
a violation of $V= 2 \sqrt{2}\approx 2.82.$ Clearly, if we add a
sixth qubit as shown in Fig.~\ref{clusterbild}(e), but do not change
the Bell operator, the violation is unchanged. The point is that, if
only stabilizing operators are considered, the graph state in
Fig.~\ref{clusterbild}(e) leads only to a violation of $5/2 = 2.5$
\cite{CGR07}, showing that the method presented here can also bring
improvements for six-qubit states.

%%%%%%%%%%%%%%%%%%%%%%%%%%%%%%%%%%%%%%%%%%%%%%%%%%%%%%%%%%%%%%%%%%%%%

%Fig. 3

%%%%%%%%%%%%%%%%%%%%%%%%%%%%%%%%%%%%%%%%%%%%%%%%%%%%%%%%%%%%%%%%%%%%%

\begin{figure}[t!!]
\centerline{\includegraphics[width=0.9\columnwidth]{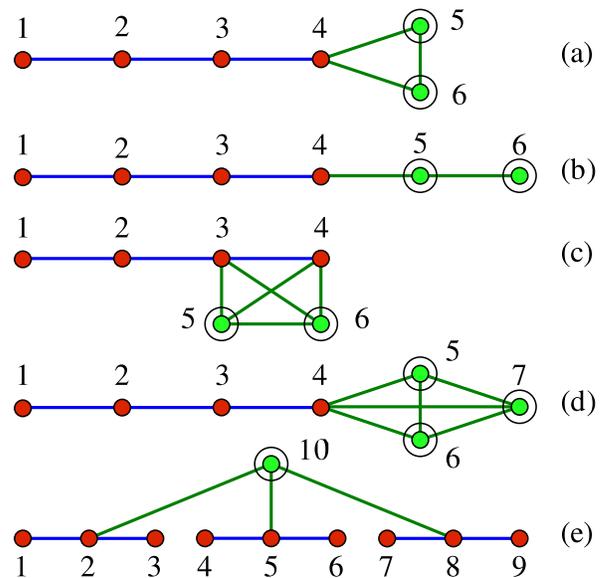}}
\caption{\label{sechsbild} (Color online) Different possibilities
for deriving Bell inequalities for graph states with six or more
qubits. See Sec.~\ref{Sec5} for details.}
\end{figure}

%%%%%%%%%%%%%%%%%%%%%%%%%%%%%%%%%%%%%%%%%%%%%%%%%%%%%%%%%%%%%%%%%%%%%

\section{The generalized method}
\label{Sec5}

%%%%%%%%%%%%%%%%%%%%%%%%%%%%%%%%%%%%%%%%%%%%%%%%%%%%%%%%%%%%%%%%%%%%%

Getting back to the derivation of the first inequality in
Eqs.~(\ref{crucial}) and (\ref{trick1}), we note that we can extend
said derivation in two directions. First, in Eq.~(\ref{crucial}) it
is not necessary that the Bell operator $\tilde{\BB}_{N-1}$
originate from ${\BB}_{N-1}$ via some transformation. In principle,
we can choose $\tilde{\BB}_{N-1}$ independently of ${\BB}_{N-1}.$
Second, in Eqs.~(\ref{trick1}), to obtain a bound for LHV models, we
used an argument similar to the one in the
Clauser-Horne-Shimony-Holt inequality \cite{CHSH69}. Here, we can
use more general inequalities, such as the
Mermin-Ardehali-Belinskii-Klyshko inequalities
\cite{BelinskiiKlyshko93}.

%%%%%%%%%%%%%%%%%%%%%%%%%%%%%%%%%%%%%%%%%%%%%%%%%%%%%%%%%%%%%%%%%%%%%

\subsection{Linear cluster states}

%%%%%%%%%%%%%%%%%%%%%%%%%%%%%%%%%%%%%%%%%%%%%%%%%%%%%%%%%%%%%%%%%%%%%

Let us consider the situation in Fig.~\ref{sechsbild}(a), where two
connected qubits are added to a four-qubit cluster state. For the
four-qubit cluster state, we consider two possible Bell operators:
first $\BB_1^{\rm (LC4)}$ from Eq.~(\ref{sasa1}), and further
$\BB_3^{\rm (LC4)}=(\openone + g_1)g_2 (\openone+g_3)$, which also
has a bound for LHV models of 2 \cite{SASA05, Cabello05, TGB06,
CGR07}. For the graph state in Fig.~\ref{sechsbild}(a), we consider
the six-qubit operator
\begin{subequations}
\begin{align}
\BB^{\rm (3a)} &= \BB_1^{\rm (LC4)} (\openone + g_5 g_6) +
\BB_3^{\rm (LC4)}(g_5 + g_6) \label{extarde1}
\\
&= [\BB_1^{\rm (LC4)}]_{[1,4]} (ZZ - XX) + [\tilde{\BB}_3^{\rm
(LC4)}]_{[1,4]} (XZ + ZX),
\end{align}
\end{subequations}
where $[\BB_1^{\rm (LC4)}]_{[1,4]}$ denotes the restriction of
$\BB_1^{\rm (LC4)}$ on the first four qubits, and
$\tilde{\BB}_3^{\rm (LC4)}$ is the Bell operator obtained from
$\BB_3^{\rm (LC4)}$ by exchanging $Z \leftrightarrow \openone$ on
the fourth qubit.

The point of the ansatz of Eq.~(\ref{extarde1}) is that the
inequality can be considered as a three-body Mermin inequality,
where $[\BB_1^{\rm (LC4)}]_{[1,4]}$ and $[\tilde{\BB_1}^{\rm
(LC4)}]_{[1,4]}$ take the role of the observables on the first
party. Since the maximum values for LHV models for $[\BB_1^{\rm
(LC4)}]_{[1,4]}$ and $[\tilde{\BB_1}^{\rm (LC4)}]_{[1,4]}$ are 2,
the bound for the total Bell operator $\BB^{\rm (3a)}$ is 4, and the
graph state in Fig.~\ref{sechsbild}(a) violates local realism by a
factor of 4.

After a local complementation on the fifth qubit, the state in
Fig.~\ref{sechsbild}(a) is equivalent to the six-qubit linear
cluster state in Fig.~\ref{sechsbild}(b). The transformed Bell
operator reads
\begin{equation} \label{extarde2} \BB^{\rm (LC6)} =
(\openone + g_1) g_2 (\openone + g_3)(\openone + g_4) g_5 (\openone
+ g_6),
\end{equation}
and has only stabilizing operators \cite{CGR07}. In fact, it can be
seen as a product of two three-qubit Mermin operators.

%%%%%%%%%%%%%%%%%%%%%%%%%%%%%%%%%%%%%%%%%%%%%%%%%%%%%%%%%%%%%%%%%%%%%

Interestingly, this method may be iterated as follows. If we
consider the seven-qubit linear cluster state, we can write down two
Bell inequalities leading to a violation of 4: First, we use the
six-qubit Bell operator from Eq.~(\ref{extarde2}), i.e., $\BB_1^{\rm
(LC7)} = (\openone + g_1) g_2 (\openone + g_3)(\openone + g_4) g_5
(\openone + g_6).$ Then, we can use its product with the generator
$g_7,$ i.e., $\BB_2^{\rm (LC7)} = (\openone + g_1) g_2 (\openone +
g_3)(\openone + g_4) g_5 (\openone + g_6)g_7.$ Taking both Bell
operators, we can use again the trick used in Fig.~3(a) to obtain a
Bell inequality for a nine-qubit graph state. After local
complementation, this leads to a Bell inequality for the nine-qubit
linear cluster state, which is the product of three three-qubit
Mermin inequalities.

Finally, it should be noted that it is not trivial that the product
of two three-qubit Mermin operators is a Bell operator in which the
total violation is the product of the two violations, since the
three-qubit graphs are connected and not independent. To give an
example, if we consider the Bell operator in Eq.~(\ref{extarde2})
and connect the first and sixth qubit (in order to form a six-qubit
ring cluster state), it will {\it not} lead to a violation by a
factor of 4, but to a violation by only a factor of 2. In general,
one can prove (along the lines of the proof of Lemma 3 in
Ref.~\cite{GTHB05}) that, if $|\meanlhv{\BB_1}| \leq C_1$ and
$|\meanlhv{\BB_2}| \leq C_2$ are two Mermin-GHZ-type inequalities on
two graphs $G_1$ and $G_2$ and these graphs become connected by a
single edge, then the Bell operators can be multiplied and the
threshold for LHV models is the product of $C_1$ and $C_2.$

%%%%%%%%%%%%%%%%%%%%%%%%%%%%%%%%%%%%%%%%%%%%%%%%%%%%%%%%%%%%%%%%%%%%%

\subsection{Six-qubit Y state}

%%%%%%%%%%%%%%%%%%%%%%%%%%%%%%%%%%%%%%%%%%%%%%%%%%%%%%%%%%%%%%%%%%%%%

Another way to obtain a generalized Ardehali inequality is shown in
Fig.~\ref{sechsbild}(c). Here, we take the Bell operators
$\BB_2^{\rm (LC4)} = (\openone+g_1)g_2 (g_3 + g_4)$ and $\BB_4^{\rm
(LC4)} = (\openone+g_1)g_2 (\openone + g_3 g_4)$, which are both
bounded by 2 for LHV models, and consider
\begin{subequations}
\begin{align}
\BB^{\rm (3c)} &= \BB_2^{\rm (LC4)} (\openone + g_5 g_6) +
\BB_4^{\rm (LC4)}(g_5 + g_6) \label{extarde3}
\\
&= [\BB_2^{\rm (LC4)}]_{[1,4]} (ZZ - XX) + [\tilde{\BB}_4^{\rm
(LC4)}]_{[1,4]} (XZ + ZX),
\end{align}
\end{subequations}
where $[\tilde{\BB}_4^{\rm (LC4)}]_{[1,4]} = ZX \openone Z
+YY\openone Z -ZYYX + Y X Y X$ is a transformed version of
$\BB_4^{\rm (LC4)}.$ With the same argumentation as before, we have
for LHV models $\BB^{\rm (3c)}\leq 4,$ so the state in
Fig.~\ref{sechsbild}(c) has a violation by a factor of 4.

This state can be transformed to the six-qubit Y state of
Fig.~\ref{clusterbild}(d) by local complementation on the fourth
qubit. Transforming the Bell operator $\BB^{\rm (3c)}$ accordingly,
one arrives at the Bell operator
\begin{equation} \BB^{\rm (Y6)}
=(\openone + g_1) g_2 (\openone+g_3)g_4 (\openone+g_5)
(\openone+g_6),
\end{equation}
which contains only stabilizing operators \cite{CGR07}. It leads to
a violation of 4, even if the replacement $A=(Z+Y)/\sqrt{2}$ and
$B=(Z-Y)/\sqrt{2}$ is {\it not} done. Again, one can iterate this
procedure in order to obtain Bell inequalities with a high amount of
violation for tree graph states (i.e., graph states associated to
graphs which do not contain any closed loop).

%%%%%%%%%%%%%%%%%%%%%%%%%%%%%%%%%%%%%%%%%%%%%%%%%%%%%%%%%%%%%%%%%%%%%

\subsection{Further extensions}

%%%%%%%%%%%%%%%%%%%%%%%%%%%%%%%%%%%%%%%%%%%%%%%%%%%%%%%%%%%%%%%%%%%%%

Let us briefly mention further extensions and applications of the
presented method. First, we can easily derive new Bell inequalities
with a high amount of violation for the case in which three or more
connected qubits are added [see Fig.~\ref{sechsbild}(d)]. Then,
however, one often does have to make the replacement
$A=(Z+Y)/\sqrt{2}$ and $B=(Z-Y)/\sqrt{2}$ in order to obtain a Bell
inequality with a high amount of violation.

%%%%%%%%%%%%%%%%%%%%%%%%%%%%%%%%%%%%%%%%%%%%%%%%%%%%%%%%%%%%%%%%%%%%%

%\subsection{Three-qubit graphs connected with a qubit}

%%%%%%%%%%%%%%%%%%%%%%%%%%%%%%%%%%%%%%%%%%%%%%%%%%%%%%%%%%%%%%%%%%%%%

Further, the presented methods can be used if several graphs become
connected. As shown in Fig.~\ref{sechsbild}(e) one may take three
three-qubit graphs, and connect them via one additional qubit.
Proper Bell operators for each of the three qubit graphs are $\BB_1
= (\openone + g_1) g_2 (\openone + g_3),$ etc., and a Bell operator
for the three disconnected graphs is $\BB = \BB_1 \BB_2 \BB_3,$
leading to a total violation by a factor of 8. For the ten-qubit
graph shown in Fig.~\ref{sechsbild}(e), we can directly write an
Ardehali-like Bell inequality, with a violation of $8\sqrt{2}.$
Similarly, one can build two- or three-dimensional structures out of
smaller graphs, and obtain for them Bell inequalities with a high
amount of violation. Such larger structures are of great interest,
since one-dimensional structures (such as the linear cluster state)
are typically not universal resources for measurement-based quantum
computation \cite{VMDB06, GE07}. Further results on this problem
will be given elsewhere.

%%%%%%%%%%%%%%%%%%%%%%%%%%%%%%%%%%%%%%%%%%%%%%%%%%%%%%%%%%%%%%%%%%%%%

\section{Conclusion}
\label{Sec6}

%%%%%%%%%%%%%%%%%%%%%%%%%%%%%%%%%%%%%%%%%%%%%%%%%%%%%%%%%%%%%%%%%%%%%

In conclusion, we have introduced a general method for deriving
Ardehali-type Bell inequalities for graph states. We have applied
the method to a variety of different graphs and showed that the
obtained inequalities often lead to a higher violation of local
realism than stabilizer-type Bell inequalities. We have generalized
the method and also obtained Bell inequalities with a high amount of
violation for tree graph states. Deriving Bell inequalities with a
high amount of violation for graph states associated to general
two-dimensional lattices remains an interesting and open problem for
further study.

%%%%%%%%%%%%%%%%%%%%%%%%%%%%%%%%%%%%%%%%%%%%%%%%%%%%%%%%%%%%%%%%%%%%%

\section*{Acknowledgments}

%%%%%%%%%%%%%%%%%%%%%%%%%%%%%%%%%%%%%%%%%%%%%%%%%%%%%%%%%%%%%%%%%%%%%

We thank H.J. Briegel for useful discussions and acknowledge support
from the EU (OLAQUI, SCALA, and QICS), the FWF, the Junta de
Andaluc\'{\i}a Excellence Project No. P06-FQM-02243 and the Spanish
MEC Project No. FIS2005-07689.

%%%%%%%%%%%%%%%%%%%%%%%%%%%%%%%%%%%%%%%%%%%%%%%%%%%%%%%%%%%%%%%%%%%%%

%%%%%%%%%%%%%%%%%%%%%%%%%%%%%%%%%%%%%%%%%%%%%%%%%%%%%%%%%%%%%%%%%%%%%


\begin{thebibliography}{14}
\bibitem{graphremark}
Since we are working now in the original graph state formalism, we
have to make the replacement $X \mapsto Y; Y \mapsto -X$ to obtain
$\tilde{\BB}$ from $\BB.$
\end{thebibliography}
\end{document}